\documentclass[aps,prl,twocolumn,groupedaddress,showpacs,showkeys,floatfix]{revtex4-1}
\usepackage{graphicx}
\usepackage{colordvi}
\usepackage{color}
\usepackage{amsmath}
\usepackage{amssymb}
 \usepackage{comment}
\newcommand{\ve}[1]{\mbox{\boldmath $#1$}}

\begin{document}
\title{Evolution of Genetic Redundancy :\\ The Relevance of Complexity in Genotype-Phenotype Mapping }


\author{Nen Saito}
\email[]{saito@complex.c.u-tokyo.ac.jp, Tel.: +81-3-5454-6732}
\author{Shuji Ishihara}
\author{Kunihiko Kaneko}
\affiliation{
Graduate School of Arts and Sciences
The University of Tokyo 3-8-1 Komaba, Meguro-ku, Tokyo 153-8902, Japan}
\date{\today}
\begin{abstract}
Genetic redundancy is ubiquitous and can be found in any organism.
However, it has been argued that genetic redundancy reduces
total population fitness, and therefore, redundancy is unlikely to evolve.
In this letter, we study an evolutionary model with high-dimensional
genotype-phenotype mapping (GPM) to
investigate the relevance of complexity in GPM to the evolution of genetic
redundancy.
By applying the replica method to deal with quenched randomness,
the redundancy dependence of the fitness is analytically obtained, which
demonstrates that genetic redundancy can indeed evolve, provided that the
GPM is complex.
Our result provides a novel insight into how genetic redundancy evolves.
\end{abstract}
\maketitle

All living organisms are under selection pressures, which act on their phenotypes, while only their genotypes are heritable. 
Therefore , the connection between genotype and phenotype, referred to as
genotype-phenotype mapping (GPM), is indispensable to fully
understand evolutionary processes. However, GPM is generally
complex and stochastic. Many phenotypic traits are now known to be a
result of complex processes involving interactions between many
proteins, RNAs, and genes. For example, developmental processes are
largely regulated by transcriptional networks that are modeled by
high-dimensional and non-linear equations~\cite{jaeger2004dynamic,fujimoto2008network}. Since such inherent
complexity in GPM obscures which genotype is associated with a high-fitness phenotype, this complexity can have an impact on evolution. However,
the relevance of GPM complexity to evolutionary processes has not
been fully explored as yet.

An important characteristic of GPM is genetic redundancy, i.e .,
coding of a phenotypic trait by two or more genes.
Numerous examples of genetic redundancy have been found in higher organisms~\cite{molin2000evolutionary,baugh2005synthetic} and even in microorganisms~\cite{musso2008extensive,krakauer2002redundancy}.
A classical premise of evolutionary theory is that genetic redundancy lowers fitness at the population
level, thus the redundancy would be evolutionarily suppressed~\cite{haigh1978accumulation,thomas1993thinking,lynch1993mutational}. 
For example,
studies have shown that genes that have been duplicated lose its function in one of the redundant genes~\cite{haldane1933part,fisher1935sheltering,nei1973probability,walsh1995often}.
According to this argument, genetic redundancy reduces the sensitivity of fitness to mutation, and thus deleterious mutations may not be eliminated, leading to a decrease in the total fitness of the population. In contrast, individuals without genetic redundancy are generally more susceptible to the deleterious effects of mutations, so much so that most mutants are lethal and mutants with lower fitness are effectively removed, thus maintaining the high fitness of the population.
This suggests that genetic redundancy is evolutionarily
unstable. In particular, the suppression of redundancy is
pronounced in asexual populations. Several studies have
explored the conditions that enable the evolution of genetic redundancy, but it is still not well understood~\cite{thomas1993thinking,nowak1997evolution,wagner2000role,krakauer2002redundancy,wagner2005robustness}.
Specifically, the relevance of complexity in GPM to the evolution of
genetic redundancy has not been evaluated.

In the present Letter, we take an asexual evolutionary model and study genetic redundancy to evolution, by comparing the results from simple and complex GPMs. Unlike the classical view mentioned above, we find that populations with higher genetic redundancy can have higher
fitness under complex GPMs . This preference for genetic redundancy is characteristic of the high-dimensional complex GPM and is
independent of previously reported mechanisms ~\cite{thomas1993thinking,nowak1997evolution,wagner2000role,krakauer2002redundancy,wagner2005robustness}.
Therefore, this study will provide a novel explanation for the ubiquity of genetic redundancy in biological systems.

In order to model both simple and complex GPM, we introduce an adiabatic spin system, where complex GPM is represented as a non-trivial mapping with quenched randomness.
Indeed, Sakata et al.~\cite{sakata2011replica} adopted a spin system with quenched randomness to study evolution (not of genetic redundancy) under complex GPM.

In  this model, configuration of $N$ sites of a locus is described by $\ve{g}=(g_{1},g_{2},\dots,g_{N})$, where each $g_{i}$ can take two different allelic states $g_{i}=\pm 1$.  Genotype $\ve{g}$ determines $M$ phenotypic traits, which is represented by $\ve{p}=(p_{1}, p_{2}, \dots, p_{M} )$, where each $p_{i}$ can take $p_{i}=\pm1$ corresponding to whether the $i$th trait is expressed or not.
Genetic redundancy is characterized by the parameter $\gamma \equiv N/M$.
For a given genotype $\ve{g}$, phenotype $\ve{p}$ is determined in a stochastic manner whose probability depends on stochastic GPM $P(\ve{p}|\ve{g})$.
We model this GPM by the following conditional probability,
\begin{equation}\label{eq:gpm}
P(\ve{p}|\ve{g}) =\exp( \beta  \sum_{ij}J_{ij}p_{i}g_{j} )/Z_{p}(\ve{g}),
\end{equation}
where $J_{ij}$ determines the mapping from $g_{j}$ to $p_{i}$ and generally can be a quenched random variable.
Here $\beta^{-1} \equiv T_{p}$ is the temperature for the stochasticity in GPM, which represents the strength of phenotypic fluctuation in isogenic individuals, and $Z_{p}(\ve{g})$ is the partition function of $\ve{p}$.
The phenotype fluctuates during an individual lifetime due to developmental noise or environmental variation. These phenotypic fluctuations are sufficiently  faster than the changes in genotype mediated by the evolutionary process of selection, reproduction, and mutation.
Thus, the phenotypic fluctuation can be adiabatically eliminated.
Hence, once $\ve{g}$ is obtained, one can assume that the distribution of $\ve{p}$ is uniquely determined by Eq.~(\ref{eq:gpm}).
Changes in genotype distribution on an evolutionary time-scale are dominated by mutation and selection processes under a given fitness function of the phenotype $\ve{p}$.  Instead of introducing a complex fitness landscape, we adopt a simple fitness function $\Phi(\ve{p})=\sum_{i}p_{i}$, to focus on the relevance of GPM to evolution.
By the adiabatic assumption, only effective fitness $\phi (\ve{g})$, the average of fitness $\Phi(\ve{p})$ over $\ve{p}$ under given $\ve{g}$, contributes to slower evolutionary dynamics of $\ve{g}$.
Distribution $P(\ve{g})$ of genotype $\ve{g}$ in the population at equilibrium is determined only by the effective fitness $\phi(\ve{g})$ and "genotypic temperature'' $T_{g}$, which represents the ratio of mutation rate to selection pressure.
The distribution of genotypes is approximated by the Boltzmann distribution as
\begin{equation}
P(\ve{g})=\exp( \beta'  \phi(\ve{g})  )/Z_{g},
\end{equation}
where $\beta'=T_{g}^{-1}$ and $Z_{g}$ is a partition function of $\ve{g}$.

We study two extreme cases of GPM in order to compare the evolutionary steady states under complex and simple GPMs. As an example of simple GPM, $J_{ij}$ is chosen to be $J_{ij}=J_{0}/N$ without randomness, whereas an example of complex GPM is represented by quenched random variables $J_{ij}$ drawn from $P(J_{ij})=\exp\left[-J_{ij}^{2}/2(\sigma_{J}^{2}/N)\right]/\sqrt{2\pi \sigma_{J}^{2}/N}$.

\begin{figure}
\includegraphics[width=9cm]{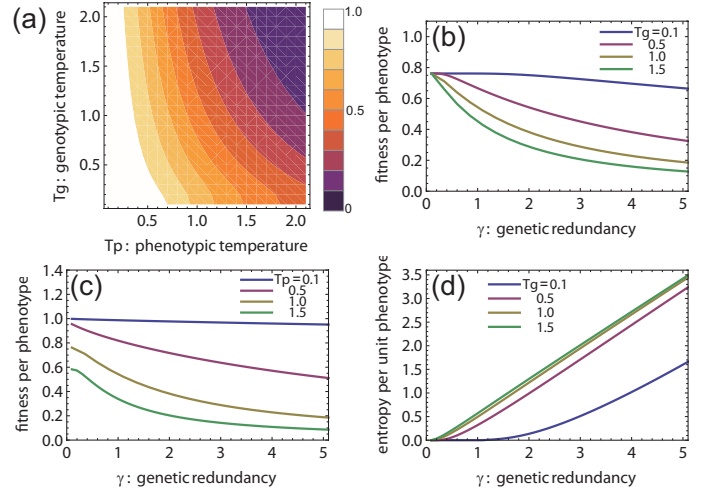}
  \caption{(COLOR ONLINE) Fitness and entropy for simple GPM.
  (a) A color map of the mean fitness per phenotype $\gamma \langle \phi \rangle$ for $\gamma =1$ is described on the $T_{p}-T_{g}$ plane.
  (b) $\gamma$ dependency of the mean fitness per phenotype $\gamma \langle \phi \rangle$ with fixed $T_{p}=1$ and $T_{g}=0.1, 0.5, 1.0$, and $1.5$.
  (c) $\gamma$ dependency of the mean fitness per phenotype $\gamma \langle \phi \rangle$ with fixed $T_{g}=1$ and $T_{p}=0.1, 0.5, 1.0$, and $1.5$.
  (d) $\gamma$ dependency of the mean entropy per phenotype $\gamma s$ with fixed $T_{p}=1$ and $T_{g}=0.1, 0.5, 1.0$, and $1.5$.
In (a)-(d), the parameter $\sigma_{J_{0}}=1$ is used.
}
\label{fig:m1}
\end{figure}

First, we consider the behavior of simple GPM $J_{ij}=J_{0}/N$.
In this case, the explicit form of effective fitness $\phi (\ve{g})$ can be calculated as $\phi(\ve{g})\equiv \mbox{Tr}_{\ve{p}}P(\ve{p}|\ve{g})\Phi(\ve{p})=M \tanh (\beta J_{0} \sum_{j}^{N}g_{j}/N)$.
For sufficiently large $N$, $Z_{g}$ is thus obtained as
\begin{eqnarray}
Z_{g} \simeq
\exp \left[ N  \left(\frac{-1+m}{2}\ln \frac{1-m}{2} -\frac{1+m}{2}\ln\frac{1+m}{2}      \right. \right. \\
\left. \left. + \frac{\beta'}{\gamma} \tanh \beta J_{0}m \right)\right],
\end{eqnarray}
where $m$ is obtained from the following saddle point equation.
\begin{equation}\label{eq:speqSGPM}
\frac{1}{2}\ln \frac{1-m}{2} -\frac{1}{2}\ln\frac{1+m}{2}     +\frac{\beta'}{\gamma} \frac{\beta J}{(\cosh \beta J_{0}m)^{2}}=0.
\end{equation}
The mean fitness per unit phenotype $\gamma \langle \phi \rangle$ is given by $\gamma \langle \phi \rangle =\frac{\gamma}{N}\frac{\partial}{\partial \beta'}\ln Z_{g}=\tanh \beta J_{0} m$.
Here, we adopt the fitness per unit phenotype $\gamma \langle \phi \rangle$, rather than the fitness per locus $\langle \phi \rangle$ in order to evaluate the contribution of genetic redundancy to fitness, because $\gamma \langle \phi \rangle$ provides a natural measure to compare systems with high redundancy (large $N$) to ones with low redundancy (small $N$), each with a fixed number of phenotypes  $M$.
In Fig.~\ref{fig:m1}-(a), $T_{g}$ and $T_{p}$ dependency of $\gamma \langle \phi \rangle$ for $\gamma=1$ is represented.
This figure illustrates that $\gamma\langle \phi \rangle$ decreases by the increase of either $T_g$ or $T_p$, i.e., with the increase in mutation rate or phenotypic fluctuation.
Fig.~\ref{fig:m1}-(b) and (c) show $\gamma$ dependency of $\gamma \langle \phi \rangle$ with fixed $T_{p}$ and $T_{g}$,
showing that $\gamma \langle \phi \rangle$ decreases as $\gamma$ increases.
Since the $\gamma$ dependency of $\gamma \langle \phi \rangle$ appears in Eq.~\ref{eq:speqSGPM} with the form $\beta' /\gamma$, one may find that increasing $\gamma$ corresponds to the increase in $T_g$ and thus leads to the decrease in fitness.
This implies that more mutations tend to accumulate in the presence of redundant genes, which leads to a reduction in fitness.  This is consistent with the suggestions that genetic redundancy weakens the effectiveness of selection in eliminating deleterious mutations in a population, thus causing the decline in population mean fitness~\cite{haigh1978accumulation,thomas1993thinking,lynch1993mutational}.
As a consequence of this accumulation of mutations, genetic redundancy enhances genetic diversity. This is represented by the increase in entropy per unit phenotype $\gamma s=\gamma(\log Z_{g}-\beta' \langle \phi \rangle)$, as shown in Fig.~\ref{fig:m1}-(d).
It follows from these results that, in the case of simple GPM, genetic redundancy $\gamma$ decreases fitness per unit phenotype $\gamma \langle \phi \rangle$ and increases entropy per unit phenotype $\gamma s$.

Next, let us consider the case with a complex GPM, where $J_{ij}$ is a quenched random variable drawn from $P(J_{ij})=\exp\left[-J_{ij}^{2}/2(\sigma_{0}^{2}/N )\right]/\sqrt{2\pi (\sigma_{0}^{2}/N)}$.
The explicit form of effective fitness $\phi (\ve{g})$ can be calculated as $\phi(\ve{g})=\sum_{i}^{M}\tanh (\beta \sum_{j}^{N}J_{ij}g_{j})$.
In this case, we need to calculate the expectation value of free energy over quenched random variables, rather than that of $Z_{g}$.
To calculate a quenched average of the free energy $\left[ F(T_{g})\right]_{\ve{J}}=\int F(T_{g})  d\ve{J}P(\ve{J})$,
the replica method~\cite{mezard1987spin} is useful, where
we first calculate $[Z_{g}^{n}]_{\ve{J}}$ for an integer $n$, and then $[\log Z_{g}]_{\ve{J}}$ is obtained from
\begin{equation}
[\log Z_{g}]_{\ve{J}}=\lim_{n \to 0}\frac{[Z_{g}^{n}]_{\ve{J}}-1}{n}.
\end{equation}
After some algebra, $[Z_{g}^{n}]_{\ve{J}}$ for an integer $n$ is obtained as
\begin{equation}
[Z]^{n}
\simeq1+N  \left(  \gamma^{-1} \log \Psi (q^{\mu\mu'})+ S(q^{\mu \mu'},\omega^{\mu\mu'})\right) ,
\end{equation}
where $q^{\mu\mu'}$ and $\omega^{\mu \mu'}$ are the spin-glass order parameter and its conjugate. Here, $\Psi (q^{\mu\mu'})$ and $S(q^{\mu\mu'},\omega^{\mu \mu'}) $ are given by
\begin{eqnarray}\label{eq:S}
S(q^{\mu \mu'},\omega^{\mu\mu'})&=\frac{1}{N}\log  \mbox{Tr}_{\{\ve{g} \}}  e^{\sum_{\mu>\mu'}   \sum _{i}^{N}g_{i}^{\mu}g_{i}^{\mu'} \omega^{\mu\mu'}}\\  \nonumber
 &- \sum_{\mu>\mu'}  q^{\mu\mu '}\omega^{\mu\mu'}
\end{eqnarray}
and
\begin{eqnarray}\label{eq:Phi}
\Psi (q^{\mu\mu'}) =&     \int \Pi_{\mu} \frac{dm_{i}^{\mu} }{\sqrt{2\pi}}    \frac{\exp \left( -\frac{1}{2\sigma^{2}} ({}^{t}\ve{m_{i}} Q^{-1} \ve{m_{i}} ) + \beta'\sum_{\mu}   f_{0}(m_{i}^{\mu}) \right) }{((\sigma^{2})^{n}\det Q)^{1/2}},
\end{eqnarray}
where $\ve{m_{i}}=(m_{i}^{1}\dots m_{i}^{n})$, $f_{0}(m^{\mu}_{i})=\tanh \beta m^{\mu}_{i}$ and $Q$ is $n\times n$ matrix with diagonal elements, 1, and off-diagonal elements, $q^{\mu\mu'}$.
For sufficiently large $N$, order parameters $q^{\mu\mu'}$ and $\omega^{\mu\mu'}$ in Eq.~(\ref{eq:S}) and Eq.~(\ref{eq:Phi}) are determined by the following saddle point equation:
\begin{eqnarray}
q^{\mu\mu'} &=\langle g^{\mu}_{i}g^{\mu'}_{i}\rangle _{L}, \\
 \omega^{\mu\mu'} &=\gamma^{-1}\frac{\partial }{\partial q^{\mu\mu'}} \log  \Psi  (q^{\mu\mu'}).
 \end{eqnarray}
Here, $\langle \cdot \rangle_{L}$ indicates the average over the distribution $\exp (\sum_{\mu >\mu'} g_{i}^{\mu}g_{i}^{\mu'}\omega^{\mu\mu'}) / \mbox{Tr}_{g_{i}^{1}\dots g_{i}^{n}} \exp (\sum_{\mu >\mu'} g_{i}^{\mu}g_{i}^{\mu'}\omega^{\mu\mu'})$.

\begin{figure}
\includegraphics[width=9cm]{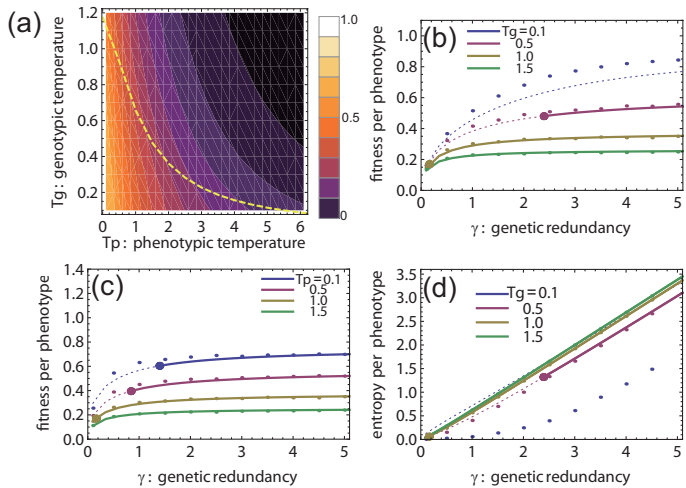}
  \caption{(COLOR ONLINE) Fitness and entropy for complex GPM.
  (a) A color map of mean fitness per phenotype $\gamma \langle \phi \rangle$ for $\gamma =1$ is described on the $T_{p}$-$T_{g}$ plane.  The yellow dashed line indicates the phase boundary between RS and RSB determined by the condition AT2(a), which is described in the Supplemental Materials.
Below the line, the system is in the RSB phase.
 (b) $\gamma$ dependency of mean fitness per phenotype $\gamma \langle \phi \rangle$ with fixed $T_{p}=1$ and $T_{g}=0.1, 0.5, 1.0$ and $1.5$.
(c)  $\gamma$ dependency of mean fitness per phenotype $\gamma \langle \phi \rangle$ with fixed $T_{g}=1$ and $T_{p}=0.1, 0.5, 1.0$ and $1.5$.
(d) $\gamma$ dependency of mean entropy per phenotype $\gamma s$ with fixed $T_{p}=1$ and $T_{g}=0.1, 0.5, 1.0$ and $1.5$.
In (a)-(d), the parameter $\sigma_{J}=1$ is used.
In (b)-(d),
thick lines indicate theoretical estimates from the RS ansatz and large circles indicate the transition point given by the AT2(a) condition.
Thin dashed lines represent invalid estimates from the RS ansatz.
In (b) and (c),
dots indicate estimates by MCS over 100 random realizations of $\ve{J}$ for $N=50$ using the Metropolis algorithm~\cite{metropolis1953equation}.
In (d), dots indicate estimates by MCS over 10 random realizations of $\ve{J}$ for $N=50$ using the multicanonical algorithm~\cite{berg1992multicanonical,berg1992new,wang2001efficient}.
}
\label{fig:1}
\end{figure}

For further calculation, we assume replica symmetry (RS) ansatz as $q^{\mu\mu'}=q$ and $\omega^{\mu\mu'}=\omega$
for all $\mu \ne \mu'$.
This ansatz provides equations for $q$ and $\omega$ as
\begin{eqnarray}\label{eq:speqRS1}
q &=\int \mathcal{D}z\left( \tanh (\sqrt{ \omega}z ) \right) ^{2}  , \\
 \omega&=\frac{\sigma ^{2}(\beta ^{'})^{2}}{\gamma}
 \int \mathcal{D}z  [f'_{0}(m)]^{2}_{\Theta},
\end{eqnarray}
where $\mathcal{D}z=  e^{ -z^{2}/2 }dz/\sqrt{2\pi} $ and $[\cdot]_{\Theta}$ indicates taking an average over the distribution $\exp (\Theta (m,z; \beta '))/\int dm \exp(\Theta (m,z; \beta ' ))$ with $\Theta (m,z; \beta ') \equiv \beta ' f_{0}(m) -  ( m+\sqrt{\sigma ^{2}q}  z  ) ^{2} /2\sigma ^{2}  (1-q)$.
For all the regions on the $T_{g}$-$T_{p}$ plane, $q$ is always positive except for when $T_{g}=\infty$ or $T_{p}=\infty$, indicating that the model has no paramagnetic phase.
By using $q$ and $\omega$ in Eq~(\ref{eq:speqRS1}), free energy per locus for the RS solution, $f_{RS}$, is calculated as
\begin{eqnarray*}
-\beta' f_{RS}&=-\frac{(1-q)\omega}{2}
+ \int \mathcal{D}z  \log 2 \cosh\left( \sqrt{ \omega}z \right)\\
&+ \int\frac{ \mathcal{D}z }{\gamma} \log \int  dm \frac{e^{\Theta(m,z;\beta')  }}{\sqrt{2\pi\sigma^{2} (1-q))}}.
\end{eqnarray*}
The RS solution becomes invalid when the de Almeida-Thouless (AT) condition~\cite{mezard1987spin,almeida1978stability} is broken, accompanied by replica symmetry breaking (RSB).
Therefore we perform Monte Carlo simulations (MCS) to estimate the fitness and entropy for the RSB phase, while theoretical estimates based on RS ansatz are used in the RS phase.
In Fig.~\ref{fig:1}-(a), the yellow dashed line shows the phase boundary between the RS and RSB phases on the $T_{g}$-$T_{p}$ plane obtained by AT conditions and indicates that RSB occurs when either $T_{g}$ or $T_{p}$ are small.
Details of the derivation of AT conditions and phase boundary are given in the Supplemental Materials.

By differentiating $-\gamma \beta' f_{RS}$ with respect to $\beta'$, the mean fitness per unit phenotype of the RS solution, $\gamma \langle \phi \rangle$, is obtained as
\begin{equation}\label{eq:meanfit}
\gamma  \langle \phi \rangle =\int \mathcal{D}z\frac{ \int  dm \tanh \beta m \  e^{\Theta (m,z;\beta')}}{ \int  dm \  e^{\Theta (m,z;\beta')}}.
\end{equation}
In the RS phase, the estimates obtained using an MCS agree with the above theoretical estimates.
A color map in Fig.~\ref{fig:1}-(a) shows $T_{p}$ and $T_{g}$ dependency of the fitness per unit phenotype $\gamma \langle \phi \rangle$ with fixed $\gamma$ and illustrates that $\gamma \langle \phi \rangle$ decreases with either $T_{p}$ or $T_{g}$ increases.
This decrease of $\gamma \langle \phi \rangle$ against $T_{p}$ and $T_{g}$ is consistent with the results of the analysis of simple GPM.
Redundancy dependence of $\gamma \langle \phi \rangle$ with fixed $T_{p}$ and $T_{g}$ is shown in Fig.~\ref{fig:1}-(b) and (c).
In contrast to the case with simple GPM, $\gamma \langle \phi \rangle$ increases with increasing $\gamma$.
This result is opposite to the classical point of view, which implies that genetic redundancy decreases fitness at evolutionary equilibrium.

Entropy per phenotype, $\gamma s$, is interpreted as the diversity of
genotypes in a population, and is calculated as $\gamma s = -\gamma \beta'
(\langle \phi \rangle+f_{RS})$ for the RS phase.
$\gamma s$ is evaluated for the RS and RSB phases numerically by using a multicanonical
Monte Carlo method~\cite{berg1992multicanonical,berg1992new,wang2001efficient},
which is often used in spin glass systems and other
fields~\cite{saito2010probability,saito2010multicanonical}.
Figure~\ref{fig:1}-(d) shows $\gamma$ dependency of $\gamma s$
with fixed $T_{p}$ and $T_{g}$. As shown in Fig.
\ref{fig:1}-(d), $\gamma s$ almost increases linearly with $\gamma$.
This increase of $\gamma s$ against $\gamma$ is similar to the case of
simple GPM, where the presence of redundant genes allows for accumulation of
mutations, and thus, an increase in the diversity of genotypes found in a population.
To conclude, in the case of complex GPM, genetic redundancy
$\gamma$ increases both fitness per unit phenotype, $\gamma \langle
\phi \rangle$, and entropy per unit phenotype, $\gamma s$.

In the present study, an adiabatic spin model with high-dimensional GPM is
investigated,  where both the complexity in GPM (represented by random
quenched variables $J_{ij}$) and genetic redundancy ($\gamma$) are
controllable parameters.
We compared evolutionary steady states between for simple and complex GPM cases.
In the case of simple GPM, fitness per unit phenotype $\gamma \langle
\phi \rangle$ decreases as $\gamma$ increases (Fig.~\ref{fig:m1}-(b)
and (c)), indicating that  redundancy should be suppressed by
selection pressure, which is consistent with the classical view~\cite{haigh1978accumulation,thomas1993thinking,lynch1993mutational}.
As is shown in Fig.~\ref{fig:m1}-(d), the decrease in redundancy
reduces entropy per unit phenotype $\gamma s$.
Since less genetic diversity in the population hinders accessibility to a novel genotype and suppresses evolvability,
selection pressure toward less genetic redundancy for simple GPM will lead a population to an evolutionary dead-end.

Remarkably, this is not true for a complex GPM. Under
complex GPM, a population with higher $\gamma$ exhibits higher fitness
$\gamma \langle \phi \rangle$, as shown in Fig.~\ref{fig:1}-(b) and
(c). Therefore, genetic redundancy, i.e., a system with higher $\gamma = N/M$
can evolve, which is contrary to the classical view of genetic redundancy.
A possible explanation is as follows:
Larger $N$ provide greater degrees of freedom in $\ve{g}$ to optimize the fitness, and thus enables realization of $\ve{g}$ that provides higher fitness than the highest fitness found in systems with smaller $N$.
At the same time, a system with smaller $M$
decreases the variety of connections from $g_{i}$ to $p_{i}$, and diminishes
frustrations in $g_{i}$ to optimize the fitness. Therefore a system with larger $N$ and
smaller $M$ can provide higher fitness.

As is shown in Fig.~\ref{fig:1}-(d), larger $\gamma$ accompanies
increases in entropy per unit phenotype $\gamma s$, and thus, in growth in genetic
diversity.
This allows a population to access a variety of novel genotypes,
which could accelerate the emergence of evolutionary innovations.
Thus, a population with complex GPM can have evolvability.
It is interesting to point out that it has been suggested that genetic diversity
enables a population to rapidly respond to large environmental
changes~\cite{wagner2005robustness,landry2007genetic,hayden2011cryptic}, thus 
our results here also suggest that a population with complex GPM could have potential ability to adapt to new environments.

The model with complex GPM exhibits RS/RSB transition.
Interestingly, such a transition was also reported in a different model by Sakata et al~\cite{sakata2011replica}, which claims that the RSB phase is biologically unfavorable.
Similar to their study, we also demonstrate that phenotypic
fluctuations suppress the appearance of the RSB phase. This RS/RSB
transition can appear in many evolutionary models with complex GPM,
and a better understanding of the biological significance of this transition
is worthy of further investigation.

In summary, we have demonstrated how complexity in GPM promotes
evolution of genetic redundancy. The mechanism is general, and
the recombination process in a sexual population does not weaken the
proposed preferences in genetic redundancy. Selection processes with
complex GPM make gene duplication preferable, which can further
enhance the potential of emerging novelty in evolution.

We would like to acknowledge helpful comments made by Koji Hukushima, and discussions with Ayaka Sakata and Tomoyuki Obuchi.
This work was supported by a Grant-in-Aid for Scientific Research (No. 21120004) on Innovative Areas ”Neural creativity for communication (No. 4103)”of MEXT, Japan.
\bibliographystyle{unsrt}


\end{document}